\documentclass{sigchi-ext}
\usepackage[T1]{fontenc}
\usepackage{textcomp}
\usepackage[scaled=.92]{helvet} 
\usepackage{graphicx} 
\usepackage{balance}  
\usepackage{booktabs} 
\usepackage{ccicons}  
\usepackage{ragged2e} 

\copyrightinfo{
  Permission to make digital or hard copies of part or all of this work for
personal or classroom use is granted without fee provided that copies are
not made or distributed for profit or commercial advantage and that
copies bear this notice and the full citation on the first page. Copyrights
for third-party components of this work must be honored. For all other
uses, contact the Owner/Author. Copyright is held by the
owner/author(s). 
}

\title{Assembly in Populations of Social~Networks}

\numberofauthors{1}
\author{%
  \alignauthor{%
    \textbf{Abigail Z.\ Jacobs}\\
    \affaddr{University of California, Berkeley} \\
    \affaddr{Berkeley, CA 94720, USA} \\
    \affaddr{azjacobs@berkeley.edu} }}

\def\plaintitle{Assembly in Populations of Social Networks} \def\plainauthor{Abigail Z. Jacobs}
\def\plainkeywords{social networks; organizational ecology; populations of organizations; social network assembly
  }

\hypersetup{%
  pdftitle={\plaintitle}, pdfauthor={\plainauthor},
  pdfkeywords={\plainkeywords}, }

\begin{document}

\maketitle

\RaggedRight{} 

\begin{abstract}

 In-depth studies of sociotechnical systems are largely limited to single instances. 
   Network surveys are expensive, and platforms vary in important ways, from interface design, to social norms, to historical contingencies. 
With single examples, we can not in general know how much of observed network structure 
is explained by historical accidents, random noise, or meaningful social processes, 
nor can we claim that network structure predicts outcomes, such as organization success or ecosystem health. 
Here, I show how we can adopt a comparative approach for settings where we have, or can cleverly construct, multiple instances of a network to estimate the natural variability in social systems.  
The comparative approach makes previously untested theories testable.
Drawing on examples from the social networks literature, I discuss emerging directions in the study of populations of sociotechnical systems using insights from organization theory and ecology.

\end{abstract}



In the late 1960s, the ecologist E.O.\ Wilson led an experiment: 
he and collaborators 
fumigated six small islands, destroying all fauna. 
This experiment built off his 1967 text \textit{The Theory of Island Biogeography}, which emphasized the opportunity of studying \emph{islands as microcosms} of large, complex ecological systems~\cite{macarthur2001theory}. 
Islands across comparable environmental conditions act as a sensible unit of observation with well-defined boundaries: for Wilson, 
observing the cleared islands revealed the empirical similarities and variation in the assembly of new ecological communities.
Contemporaneously, sociologist Peter Blau argued 
for ``the systematic comparison of a fairly large number of organizations in order to establish relationships between their characteristics'' \cite{blau1965comparative}.
However, major comparative studies remain a challenge in both fields. Now, complex sociotechnical ecosystems characterize our homes, social lives, and workplaces, 
making this challenge as relevant as ever. With the promise of newly available sources of data, we can explore how complex, interrelated underlying social processes shape the structure of these systems.
Focusing on network structure, we are left with the same challenge confronted fifty years ago: 
with single examples of social systems, we can not in general know
the degree to which observed network structure 
is explained by historical accidents, random noise, or meaningful social processes. 
Nor can we
claim that network structure predicts outcomes, such as organization success or ecosystem health.\footnote{Borrowing from Blau: ``the comparative method, in the broadest sense of the term, underlies all scientific and scholarly theorizing'' \cite{blau1965comparative}.} 
Here we focus on settings with multiple comparable \emph{networks of {organizations}}, 
with observed membership and organizational properties and outcomes.
In this work, we highlight that in settings with multiple comparable networks of organizations, we can begin to understand how network structure impacts outcomes, and how network structure reflects organizational environment and underlying social processes.
\looseness=-1

\begin{marginfigure}[-28.5pc]
  \begin{minipage}{\marginparwidth}
    \centering
    a \includegraphics[width=0.78\marginparwidth]{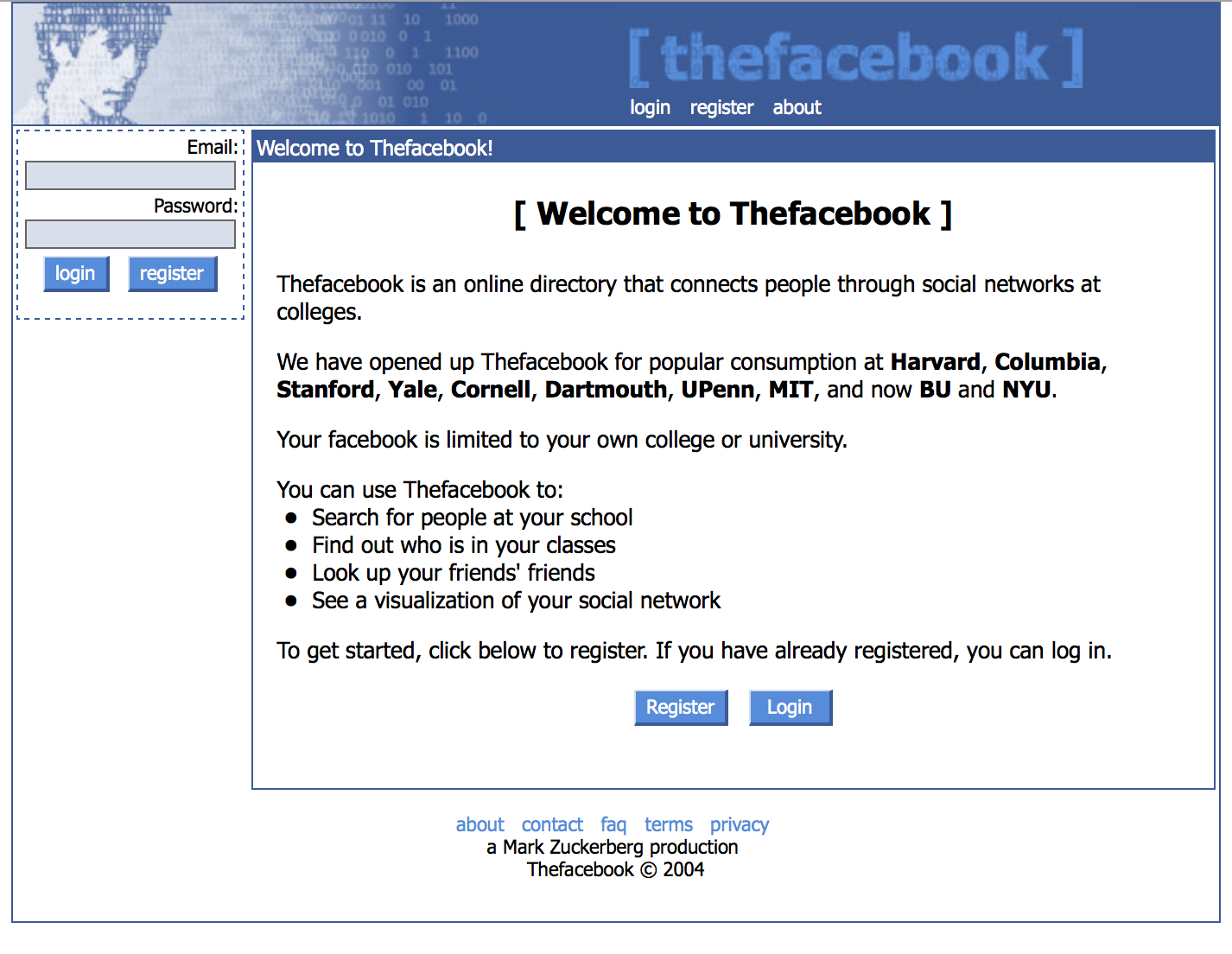}\\
    \vspace{0.2em}
    b \includegraphics[width=0.75\marginparwidth]{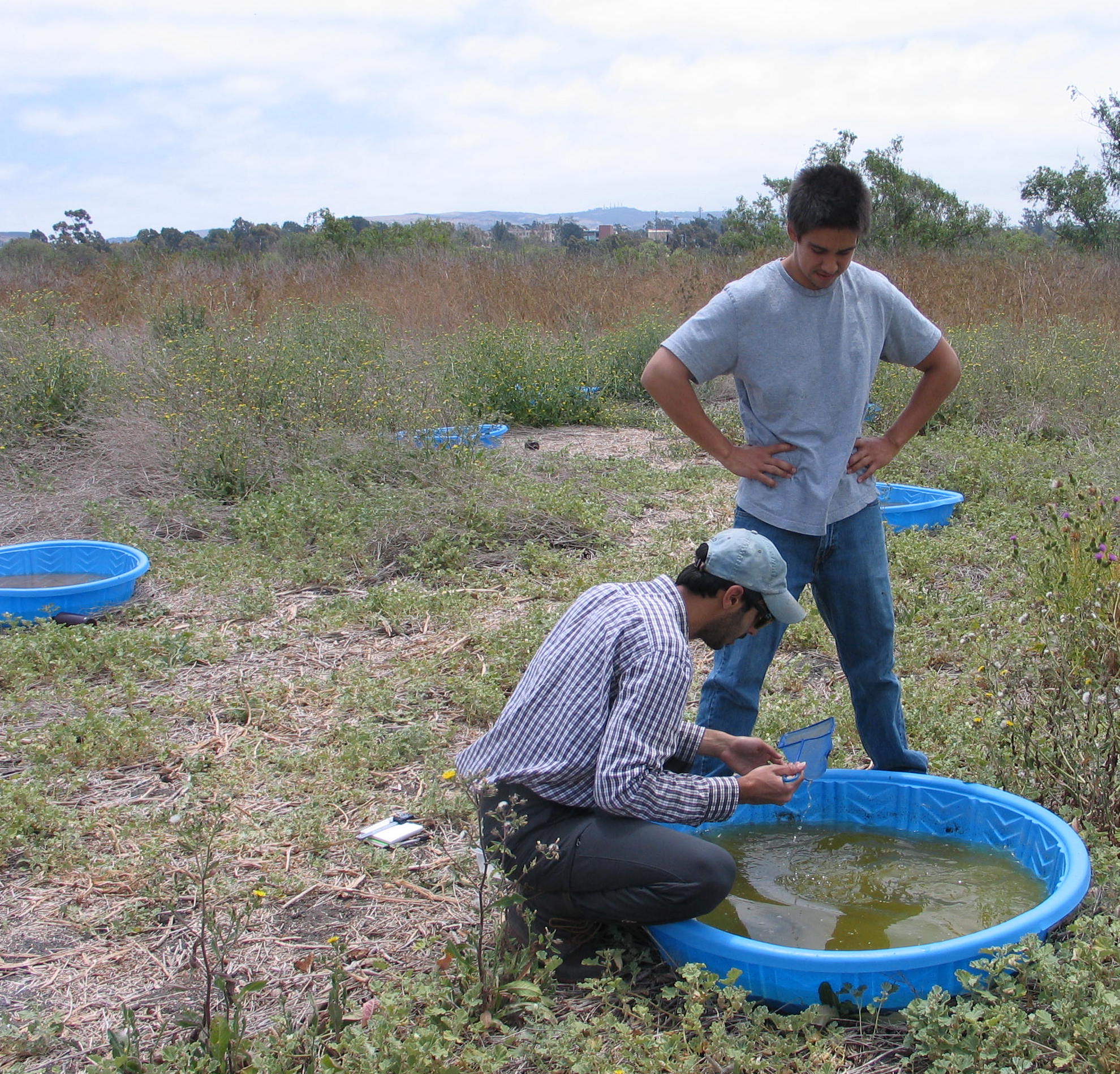}\\
    \vspace{0.2em}
    c \includegraphics[width=0.75\marginparwidth]{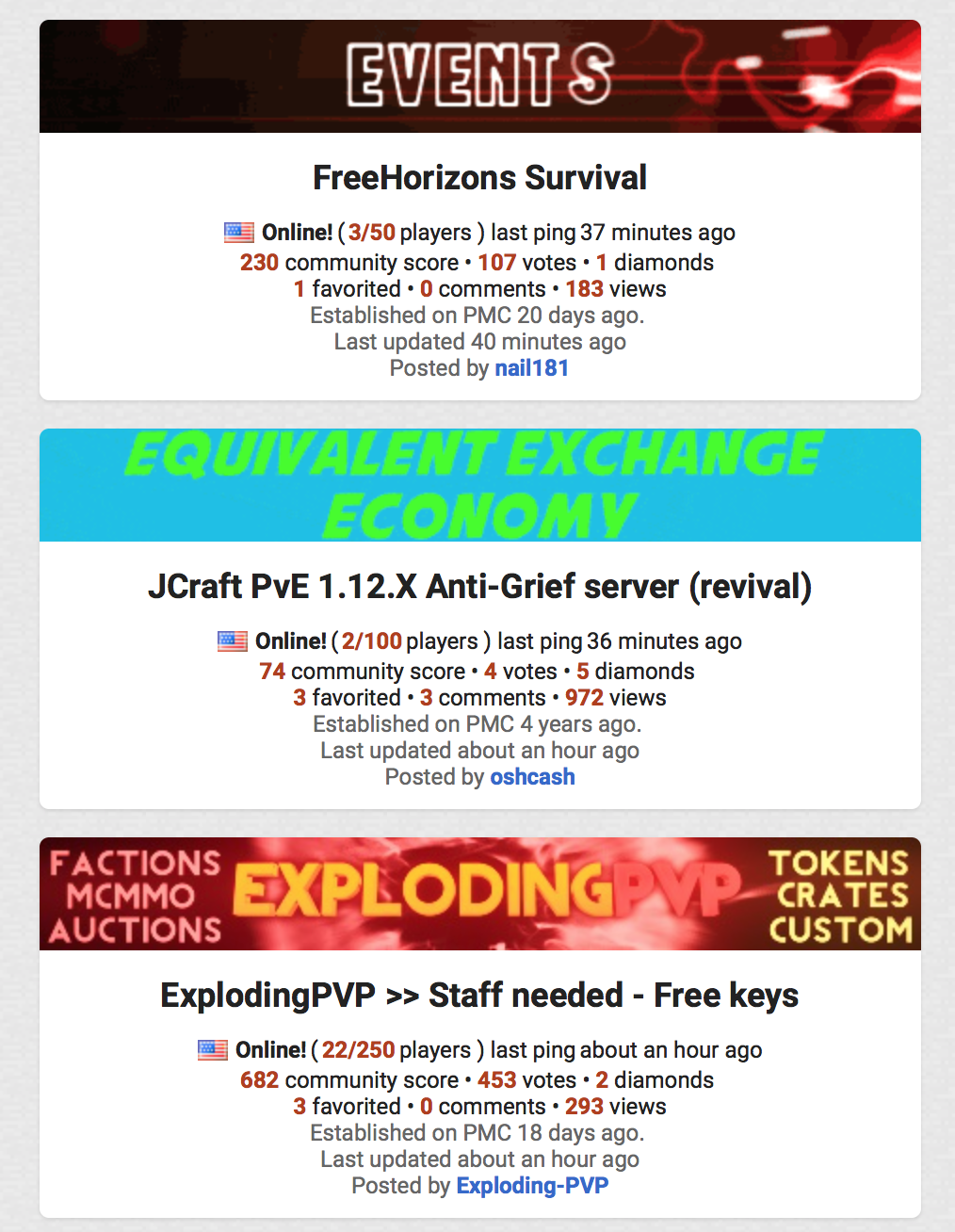}
    \\ \vspace{0.2em}
     \scriptsize{Source: {\texttt{thefacebook.com}, April 3, 2004, via the Internet Archive}; {Ashkaan Fahimipour}; \texttt{planetminecraft.com}}
    \caption{a) Facebook's iterative expansion \cite{jacobs2015assembling}; b) ecologists study assembly with temporary ``ponds,'' no fumigation necessary \cite{fahimipour2014dynamics};  c) multiple universes of Minecraft servers with different rules \cite{frey2018emergence}.
    }
  \end{minipage}
\end{marginfigure}

\paragraph{Why populations of networks?}
How do we show that how the structure of sociotechnical systems meaningfully varies with social and organizational properties?
Analysis across multiple platforms is undoubtedly useful; 
however across instances of different platforms, ruling out variation due to platform differences or historical idiosyncrasies is nontrivial. 
In contrast, among sociotechnical systems of \emph{comparable} origin or generating process, we can take a 
\textit{population-level} approach, where we can characterize variation across a population of social systems.
Uncontroversially, evidence for specific empirical social phenomena is more compelling when shown across multiple instances of comparable social systems.\footnote{For example, high school is as bad as you thought: status predicts social structure in a population of 100 high school social networks \cite{Ball2012}. 
In that setting, as here, the focus is on ``finite populations,'' drawn from comparable generating processes (100 schools, one name generator), not multiple draws from the same process (e.g., 100 observations of one school).\looseness=-1} 
Hill and Shaw persuasively argue for the population-level study of communities across a single medium, and we follow in their steps here~\cite{mako2017populationschapter}.\footnote{Paraphrasing Hill and Shaw, they highlight the benefits of such a perspective: generalizability of results across online communities; the ability to study community- or organization-level attributes and outcomes; insight into diffusion between communities, e.g., across platforms; insight into ecological dynamics, extending the organizational ecology approach to online systems; and insight into multilevel processes, merging individual-level dynamics with understanding of meso- and macro-level processes \cite{mako2017populationschapter}.}


\paragraph{Social network assembly}
We draw on the ecological notion of \emph{community assembly}, which reflects the complex, overlapping processes contributing to community formation \cite{warren2015islands}. 
These processes have direct analogs in sociotechnical systems: composition of the current community; ordering effects (which group arrives earliest may set constraints on who may join, or set norms for behavior); competition within and between systems; and natural limits on growth (due to local or global resources)  \cite{jacobs2015assembling}.
Just as islands were necessary to study community assembly, here, I show how we can adopt a comparative approach for settings where we have, or can cleverly construct, multiple instances of a network (Fig.\ 1).
This approach allows us to estimate natural variability in social systems and tease apart the social processes underlying \emph{network assembly}.

\paragraph{Boundaries of sociotechnical systems}
To empirically study populations of social systems, we must define inclusion criteria for a set of users. These criteria determine the \emph{boundary} of or within a sociotechnical system.
\marginpar{%
  \fbox{%
    \begin{minipage}{0.925\marginparwidth}
      \textbf{Example: thefacebook.com}\\
      \vspace{0.4pc} \textbf{Setting:} 
      Facebook launched iteratively to universities during 2004--05; 
      at the time the platform was
      designed for interaction within-organization, 
      with little support for cross-organization interaction.\\
      \vspace{0.4pc} \textbf{Opportunity:} Exploiting variation in timing of Facebook adoption and school start dates, Jacobs et al.\ (2015) develop a natural experiment across the first 100 universities to adopt Facebook. 
      This created a direct test of whether differences in offline social environments changed the structure of online behavior~\cite{jacobs2015assembling}.\\ 
      \vspace{0.4pc} \textbf{Results (sample):} 
Students with no in-person interaction had, on average, more male-female friendships, whereas students with significant in-person interaction had more same-gender friendships. Average number of friendships and student adoption rate increased with time on campus. Adoption rate in the university population best explained network structure, analogous to ecological community assembly~\cite{fahimipour2014dynamics}.
    \end{minipage}}\label{sec:sidebar-thefacebook} }
%
%
For example, we might consider all registered users of a platform,
or all \emph{eligible} individuals---e.g., 
all individuals with a \texttt{harvard.edu} email address in 2004 
(see panel). 
These properties can be defined exogenously from the platform, for example, platform members with a shared offline affiliation, such as an alma mater, team, or employer \cite{jacobswatts2018doesnetwork,jacobs2015assembling,phan2015experiment,zhu2014selecting}.
These properties can also be endogenous to the platform, such as Reddit communities or Facebook groups. In both settings, the onus is on the researcher to empirically or experimentally defend the degree to which interaction between communities 
affects interactions within the community.
We can then study network assembly---processes such as product adoption or the emergence of norms---by looking at interactions within our bounded populations.\looseness=-1

\paragraph{Empirical assembly and diversity}
We can exploit these boundaries to empirically explore assembly. 
For example, university affiliations reveal that Facebook adoption rates reflected students' shared geography 
\cite{jacobs2015assembling}. 
Across a population of peer production systems, 
norms are set and entrenched by early settlers of wiki governance arms~\cite{shaw2014laboratories,teblunthuis2018revisiting}.\looseness=-1
Organizational ecology also seeks to understand
the diversity and sources of heterogeneity \emph{across} and \emph{within} organizations 
\cite{baum2006eco}. 
Even among organizations of similar types---firms within the same industry, or gaming systems with shared plug-ins---recent work reveals a wide diversity across organizational forms \cite{frey2018emergence,jacobswatts2018doesnetwork} (see panel, next page).
It is impossible to fully characterize organizational diversity 
without studying systems that fail; fortunately, these settings can be cleverly empirically designed \cite{frey2018emergence,hill2013almost,teblunthuis2018revisiting}. 
To understand diversity \emph{within} organizations, 
we can characterize communities with overlap in properties, goals, and membership as belonging to the same \emph{niche} \cite{zhu2014selecting}. Within the same {niche}, communities must either compete for members or resources, or otherwise some mechanism must allow them to coexist.
In online settings, similar groups suffer competition from sharing members \cite{wang2013impact}; moreover, competition is more intense among communities with users with shared \emph{offline} affiliations \cite{zhu2014selecting}. Turmoil within a community can drive migration \cite{newell2016user}; sufficiently diverse platforms, however, can support the emergence of new niches---for example counterpublics of non-dominant but cohesive communities~\cite{jackson2015hijacking}.

\paragraph{Discussion}
The comparative perspective enables us to quantify effects of policies on platforms, how norms emerge, how organizational structure reflects their environments, and how structure reinforces desirable (or undesirable) outcomes for individuals, organizations, and systems of platforms. 
%
While experiments across networks are usually limited to virtual lab settings (e.g., \cite{centola2015spontaneous,mao2016experimental}) and observational data reveal a noisy, incomplete representation of a system \cite{gaffney2018caveat}, future work will progress by developing empirical strategies for observational data. 
Despite a plethora of enterprise communication and task-based systems and multi-community online platforms such as Reddit, Wikia and StackExchange, this area remains underdeveloped.
Organizational ecology provides novel opportunities to understand populations of sociotechnical systems. While still rare, this is a compelling area for future~research.\looseness=-1

\pagebreak
\paragraph{Acknowledgments}
The author thanks Kristen Altenburger, Aaron Clauset, Aaron Shaw, Johan Ugander, and Duncan Watts
for helpful conversations during the development of this work.

\marginpar{%
  \fbox{%
    \begin{minipage}{0.925\marginparwidth}
      \textbf{Example: Comparison of organizational networks} \\
      \vspace{0.5pc} \textbf{Setting:} 
      Within-organization communication networks for 65 U.S.-based firms, ranging in size from 
2,000 to 200,000 employees, using anonymized metadata from over two billion email exchanges.\\
      \vspace{0.5pc} \textbf{Opportunity:} In-depth studies in organization theory largely rest on evidence from single examples; empirical heterogeneity of organizations is unknown.\\
      \vspace{0.5pc} \textbf{Results (sample):} Communication is more centralized in geographically disparate organizations. Counter to predictions from organization theory, Jacobs and Watts find effectively no meaningful relationships between the performance, industry, or other firm-level attributes and organizational network structure~\cite{jacobs2017comparative,jacobswatts2018doesnetwork}. 
        \end{minipage}}\label{sec:sidebar-poporgs} }
\balance{} 
\bibliographystyle{SIGCHI-Reference-Format}

\end{document}